\documentclass[aps,prb,10pt,twocolumn]{revtex4-1}
\usepackage[colorlinks=true,urlcolor=blue,citecolor=red]{hyperref}
\usepackage{graphicx}
\usepackage{amsmath}
\usepackage{amssymb}
\newcommand{\dd}{\mathrm{d}}
\newcommand{\expval}[1]{\left \langle #1 \right\rangle}
\renewcommand{\vec}[1]{\boldsymbol{#1}}
\begin{document}
\title{Magnetic defect line in a critical Ising bath}
\author{Andrea Allais}
\affiliation{Department of Physics, Harvard University, Cambridge, MA 02138, USA}
\begin{abstract}
We compute the critical exponents associated with a magnetic line defect in the critical 3D Ising model. From the result, we deduce the anomalous dimension of the fermion operator in the $z = 1$ scaling regime of a Fermi surface coupled to a O(1) Wilson-Fisher fixed point. We introduce a generalization of the Wolff cluster algorithm to O($N$) spin systems with a background magnetic field.
\end{abstract}
\maketitle
\section{Introduction and summary of results}
The context of this work is the study of critical metals, that is, of a Fermi surface coupled to some gapless bosonic degree of freedom \cite{Lohneysen, Metlitski Sachdev 1, Metlitski Sachdev 2}. In particular, we focus on a Fermi surface coupled to a Wilson-Fisher critical boson. 

At very low energy, the boson fluctuations are damped by the Fermi surface, and the theory flows to a non-relativistic fixed point. However, there are models in which over a substantial energy range the damping effect is negligible, and the boson dynamic is simply that of a Wilson-Fisher fixed point \cite{Sachdev Sokol, Sachdev Georges, Fitzpatrick Raghu, Hartnoll Sachdev}. This regime is the focus of our work. Moreover, the Fermi velocity of the fermions flows to zero in this regime \cite{Sachdev Vojta, Fitzpatrick Raghu}, so we consider dispersionless fermions. This problem has been analyzed with perturbative methods in ref. \onlinecite{Allais Sachdev}.

Since under our assumptions the Fermions do not move and do not affect the boson, we may as well consider a single particle sitting at $\vec x = 0$. That is, we consider the hamiltonian
\begin{equation}
 H = H_{\mathrm{b}}(\pi, \phi)  - h \phi(0) \psi^{\dag} \psi\,,
\end{equation} 
where $H_{\mathrm{b}}$ is the $\mathbb{Z}_2$ invariant hamiltonian that governs the boson $\phi$, which we assume tuned to the Wilson-Fisher fixed point.  In this system, for any non-zero $h$, the fermion Green's function satisfies has following non-Fermi liquid scaling behavior\cite{Allais Sachdev}:
\begin{equation}
 G(\tau) \equiv \left\langle\psi(\tau) \psi^{\dag}(0)\right\rangle \sim \frac{e^{-\delta \mu\, \tau}}{\tau^{\eta_\psi}} \quad\text{for $\tau \to \infty$}\,,
\end{equation} 
where $\eta_\psi$ is a universal anomalous dimension, and $\delta \mu$ is a non-universal renormalization of the chemical potential. 
The main result of this work is that
\begin{align}
 & \eta_\psi = 0.187(7)\,,
\end{align}
which we obtain from a Monte Carlo simulation.

\begin{figure}
\begin{center}
\includegraphics[scale=0.5]{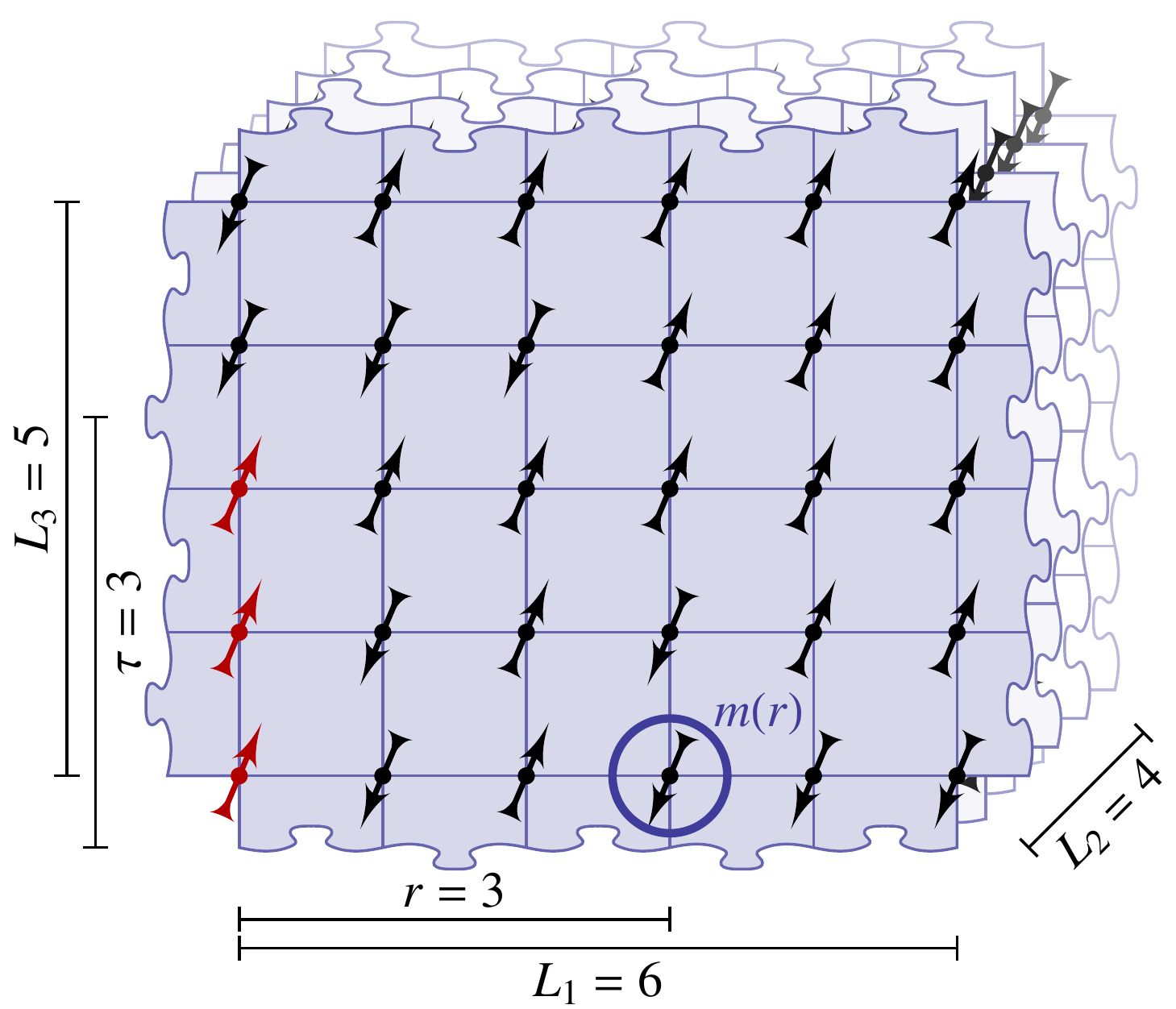}
\end{center}
\caption{\label{fig:system_geometry}Schematic representation of the geometry of the system. Classical Ising spins occupy the sites of a cubic lattice with sides $L_1$, $L_2$, $L_3$ and periodic b.c. A line of $\tau$ spins (in red) is polarized by a local magnetic field $h$. The average magnetization $m(r)$ is measured at distance $r$ from the line.}
\end{figure}

From the point of view of the boson, the problem is that of a magnetic impurity introduced at $\vec x = 0$ for a time $\tau$ and then removed:
\begin{equation}\label{eq:Green's function}
\begin{split}
 G(\tau)  = \frac{\mathrm{Tr}\left[e^{-(\beta-\tau) H_{\mathrm b}} e^{-\tau[H_{\mathrm b} - h \phi(0)]}\right]}{\mathrm{Tr}\left[e^{-\beta H_{\mathrm b}}\right]}\,.
\end{split}
\end{equation} 
A regularization of this system on the lattice is the critical 3D Ising model with a defect line created by a non-zero magnetic field (see fig. \ref{fig:system_geometry}):
\begin{align}
 Z(\tau) = e^{-F(\tau)} = \sum_{\sigma} e^{-E(\sigma)}\,,
\end{align}
with
\begin{align}
 E(\sigma) = - J \sum_{i = 1}^{3} \sum_{\vec a} \sigma_{\vec a}\sigma_{\vec a + \hat{\vec \imath}} - h \sum_{s = 0}^{\tau - 1}\left(\sigma_{s \hat{\vec 3}} - 1\right)\,,
\end{align} 
where $\vec a$ denotes the sites of a cubic lattice of sides $L_1, L_2, L_3$ and $\hat{\vec 1},\, \hat{\vec 2},\, \hat{\vec 3}$ are the generators of the lattice. We subtracted the constant $h \tau$ from the usual form of the energy so that the limit $h\to\infty$ is well defined. A similar problem of a line twist defect in the Ising model has been studied in ref. \onlinecite{Billo Pellegrini}.

According to \eqref{eq:Green's function} we have $\log G(\tau) = F(0) - F(\tau)$, and hence, in the thermodynamic limit $L_i \to \infty$, if the coupling $J$ is tuned to the critical value and for any non-zero $h$, we expect the following scaling behavior
\begin{equation} \label{eq:free energy scaling}
 F(\tau) \sim \delta \mu\, \tau + \eta_{\psi}\log \tau + O(\tau^0)\quad\text{for $\tau \to \infty$}\,.
\end{equation} 
Section \ref{sec:free energy} describes more in detail how we determine $\eta_\psi$ by fitting Monte Carlo data to this relation.

In addition to the free energy $F$, another interesting observable is the profile of the magnetization in the presence of a defect line that spans the entire system
\begin{equation}
 m(r) = \expval{\sigma_{r\hat {\vec{1}}}}_{\tau = L_3}\,,
\end{equation} 
which in the thermodynamic limit must also have a nontrivial scaling behavior
\begin{equation}\label{eq:magnetization}
 m(r) \sim \frac{c_1}{r^{\Delta}}\quad \text{for $r \to \infty$}\,,
\end{equation} 
where the scaling dimension $\Delta$ is universal. As described in section \ref{sec:magnetization profile}, by fitting Monte Carlo data to this relation we determine
\begin{align}
\Delta = 0.526(5)\,.
\end{align}
This result is broadly consistent with the hypothesis\cite{Allais Sachdev} $\Delta = (1 + \eta)/2 = 0.5182(3)$, where $\eta = 0.0364(5)$ is the anomalous dimension of the $\sigma$ operator \cite{Pellissetto}.

In section \ref{sec:simulation} we describe the details of the simulation strategy, and in particular how to include a magnetic field in the Wolff cluster algorithm for the Ising model.

\section{Green's function}
\label{sec:free energy}
It is not possible to compute the free energy $F$ directly, but it is possible to compute the free energy difference between two values of $\tau$:
\begin{equation}
 e^{- F(\tau_2) + F(\tau_1)} = \expval{\exp\left[h \sum_{s = \tau_1}^{\tau_2 - 1}\left(\sigma_{s \hat{\vec 3}} - 1\right)\right]}_{\tau = \tau_1}\,.
\end{equation} 
This quantity becomes increasingly noisy as $\tau_2 - \tau_1$ grows, so we compute it for a sequence for pairs $(\tau, \tau + 1)$. Moreover, for simplicity, we take the limit $h \to \infty$, and we have
\begin{equation}
 F(\tau + 1) - F(\tau) = - \log \expval{\theta[\sigma_{\tau\hat{\vec 3}}]}_{\tau}\,,
\end{equation} 
where $\theta$ is the Heaviside theta. 
\begin{figure}
\begin{center}
\includegraphics[scale=0.5]{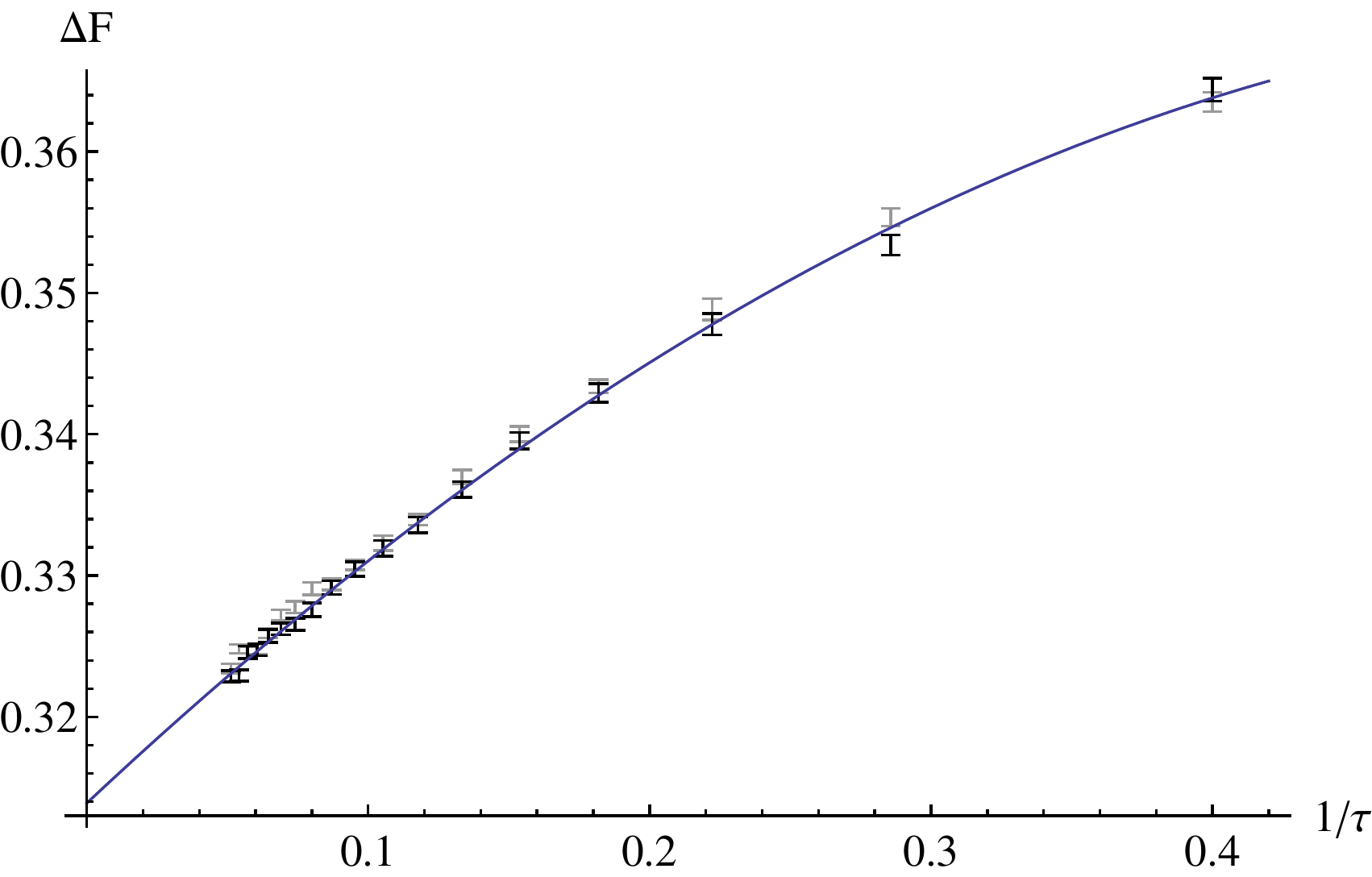}
\end{center}
\caption{\label{fig:free energy fit} The free energy difference \eqref{eq:free energy difference} for system size $256^3$ (black) and $128^3$ (light gray). The solid line is the best fit to the $256^3$ data of the form \eqref{eq:free energy difference}. $\chi^2 = 10$ with 15 degrees of freedom.}
\end{figure}
Based on the scaling form \eqref{eq:free energy scaling} we model the data as
\begin{equation}\label{eq:free energy difference}
 \Delta F(\tau) \equiv F(\tau + 1/2) - F(\tau - 1/2) = \delta\mu + \frac{\eta_\psi}{\tau} + \frac{c_3}{\tau^2}\,.
\end{equation} 
Here $\tau$ takes half integer values, and the last term is a correction to scaling, negligible at large $\tau$.

Fig.~\ref{fig:free energy fit} displays the data and the fit to this form. Every point is the average of a number of identical independent simulations. The observable is very noisy and the statistical errors we obtain employing reasonable computational resources are bigger than the finite size effects at size $128^3$ and $256^3$. Therefore, there is no use for finite-size scaling.

We find
\begin{align}
 &\delta\mu = 0.3139(4)\,,
 &&\eta_\psi = 0.187(7)\,,
 &&c_3 = -0.16(2)\,,
\end{align}
where the errors are evaluated with a statistical bootstrap of the whole data set.

\section{Magnetization profile}
\label{sec:magnetization profile}
\begin{figure}
\begin{center}
\includegraphics[scale=0.5]{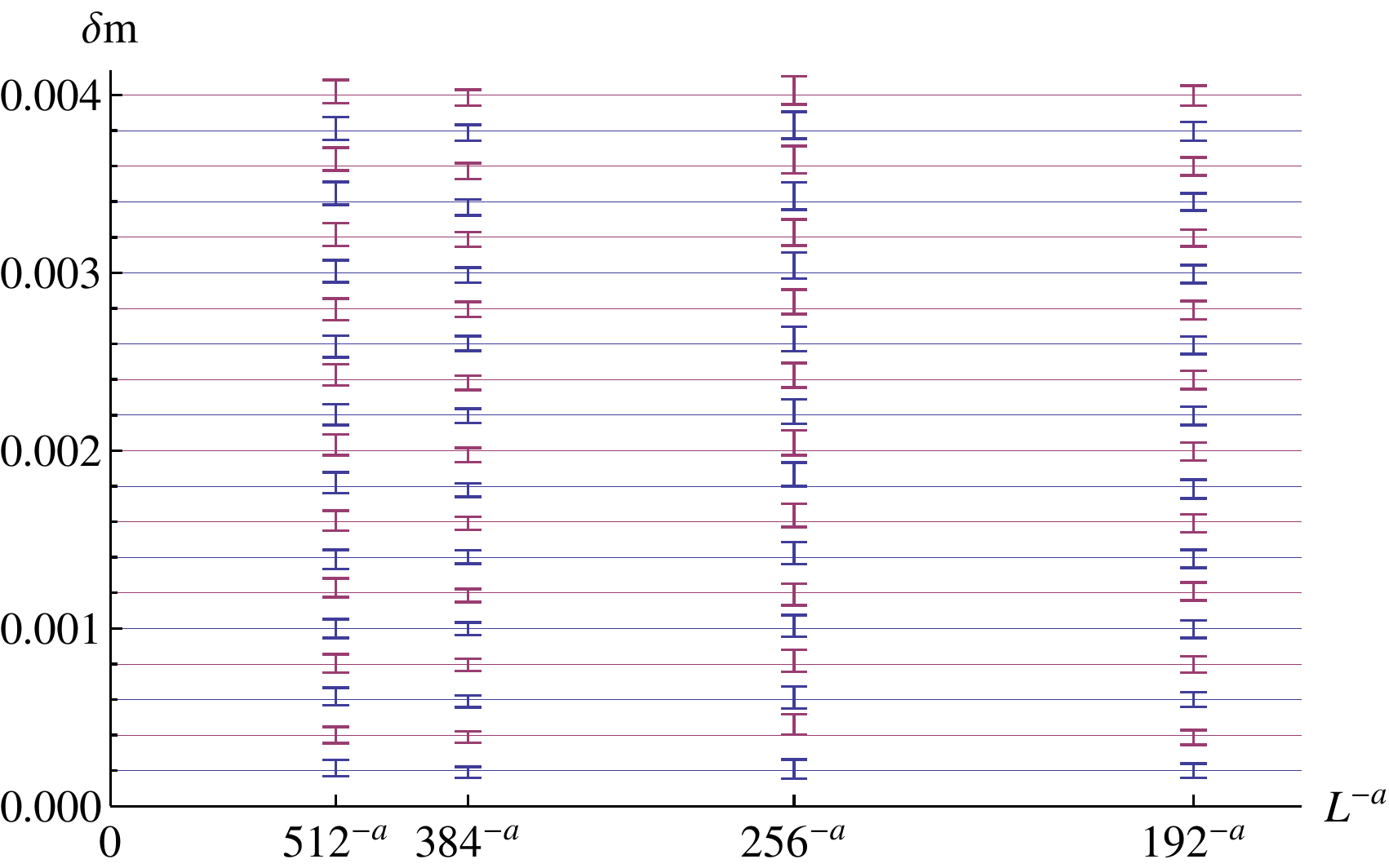}
\end{center}
\caption{\label{fig:finite size residuals} Residuals of the finite size scaling fit \eqref{eq:finite size scaling} with $a = 1.60$, from $r = 6$ at the bottom to $r = 25$ at the top. Total $\chi^2 = 18$ with 37 degrees of freedom.}
\end{figure}

We compute the profile of the magnetization \eqref{eq:magnetization} for system sizes $192^3$, $256^3$, $384^3$ and $512^3$. For every system size, we carry out a number of identical independent simulations, from which the profile is extracted, with the corresponding error. The statistical errors are sufficiently small to allow for finite size scaling. We assume the form
\begin{equation}\label{eq:finite size scaling}
 m(r, L) = m(r) + c(r) L^{-a}\,,
\end{equation} 
with $a = 1.60$. The residuals of the fit are shown in fig.~\ref{fig:finite size residuals}. We evaluate the errors on $m(r)$ with a statistical bootstrap of the whole data set.

\begin{figure}
\begin{center}
\includegraphics[scale=0.5]{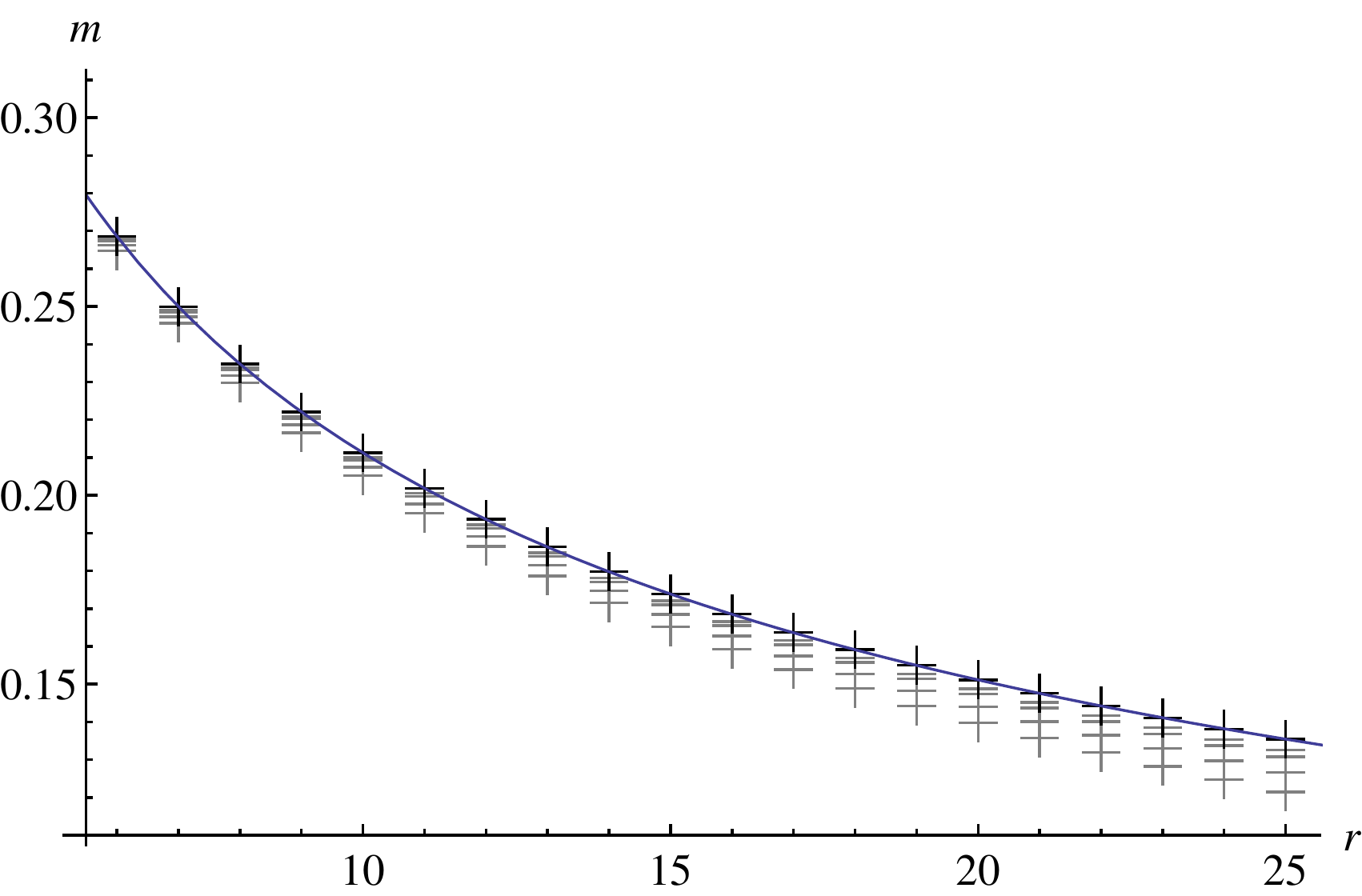}
\end{center}
\caption{\label{fig:magnetization profile} Magnetization profile $m(r)$ in the infinite system limit (black), and for system sizes $192^3$, $256^3$, $384^3$ and $512^3$ (light gray). Error bars are too small to show. The solid line is the best fit to the infinite system data of the form \eqref{eq:magnetization scaling}. $\chi^2 = 7.5$ with 15 degrees of freedom. }
\end{figure}

For the range of $r$ we can access, there are substantial corrections to scaling. Therefore we assume the form
\begin{equation}\label{eq:magnetization scaling}
 m(r) = \frac{c_1}{r^{\Delta}} + \frac{c_2}{r^{\Delta_2}}\,,\quad \Delta_2 > \Delta\,,
\end{equation} 
and we find
\begin{align}
 &\Delta = 0.526(5)\,, 
&&\Delta_2 =0.93(3)\,,\\
&c_1 = 0.80(3)\,,
&&c_2 = -0.22(3)\,.
\end{align}
Here as well errors are determined using statistical bootstrap. The data and the best fit are shown in fig.~\ref{fig:magnetization profile}.

\section{Simulation strategy}
\label{sec:simulation}
When studying critical phenomena it is crucial to use algorithms that do not suffer from critical slowdown. Luckily there exists \cite{Binder} a straightforward generalization of Wolff algorithm \cite{Wolff} that allows the simulation in presence of a background magnetic field. Let us briefly discuss it.

We want to simulate the partition function
\begin{equation}
 Z = \int \prod_{i} \dd \vec \sigma_i\ e^{\sum_{ij} J_{ij} \vec \sigma_i\cdot \vec \sigma_j + \sum_i h_i \vec u \cdot \vec \sigma_i}\,,
\end{equation} 
where $\vec \sigma_i$ are unit vectors on the nodes of a graph, and $\vec u$ is a unit vector specifying the direction of the magnetic field. In order to do so, we introduce the  partition function of an extended system which involves an extra O($N$) integration variable $\vec \rho$:
\begin{equation}
 Z_1 = \int \dd \vec \rho \prod_{i} \dd \vec \sigma_i\ e^{\sum_{ij} J_{ij} \vec \sigma_i\cdot \vec \sigma_j + \sum_i h_i \vec \rho \cdot \vec \sigma_i}\,.
\end{equation} 
This is the partition function of an O($N$) model with no magnetic field, defined on a different graph in which the $\vec \rho$ spin is coupled to every other spin $\vec \sigma_i$ with a coupling $h_i$. This partition function is suitable for simulation with the Wolff algorithm. We also introduce an orthogonal matrix $R(\vec \rho|\vec u)$ such that
\begin{equation}
 R(\vec \rho|\vec u) \vec \rho = \vec u\,,
\end{equation} 
for example the Householder reflection
\begin{align}
 &R(\vec \rho|\vec u) = 1 - 2 \frac{(\vec u - \vec \rho)(\vec u - \vec \rho)^T}{|\vec u - \vec \rho|^2}\,.
\end{align}

It is easy to show that, for any observable $F[\vec \sigma]$, we have
\begin{equation}
 \expval{F[\vec \sigma]}_{Z} = \expval{F[R(\vec \rho) \vec \sigma]}_{Z_1}\,.
\end{equation} 

In our case, we deal with the $\mathrm{O}(1) = \mathbb Z_2$ (Ising) model. The magnetic field is nonzero only on a line of sites. For simplicity, we also go to the extreme limit of an infinite magnetic field, so that the spins of this line are fully polarized. In the extended model, all these spins are locked to the spin $\rho$ by the infinite coupling $h_i$, and hence also locked together. Therefore, the simulation boils down to applying the Wolff algorithm to the original system with all the couplings $J_{ij}$ between spins on the line set to infinity. When computing an observable, all the spins involved are first multiplied by the common sign of the spins belonging to the line. This is the equivalent of the multiplication by $R(\vec \rho|\vec u)$ in this simpler setting.

\section{Acknowledgements}
We thank S. Sachdev for suggesting to look at this problem, and F. Parisen Toldin for pointing out ref.~\onlinecite{Binder}. This research was supported by the NSF under Grant DMR-1103860, and by the Templeton foundation.


\thebibliography{unsrt}
\bibitem{Lohneysen} H. v. L\"ohneysen, A. Rosch, M. Vojta, and P. W\"olfle, Fermi-liquid instabilities at magnetic quantum phase transitions, \href{http://journals.aps.org/rmp/abstract/10.1103/RevModPhys.79.1015}{Rev. Mod. Phys. \textbf{79}, 1015}.

\bibitem{Metlitski Sachdev 1} M. A. Metlitski and S. Sachdev, Quantum phase transitions of metals in two spatial dimensions. I. Ising-nematic order, \href{http://journals.aps.org/prb/abstract/10.1103/PhysRevB.82.075127}{Phys. Rev. B \textbf{82}, 075127}.

\bibitem{Metlitski Sachdev 2} M. A. Metlitski and S. Sachdev, Quantum phase transitions of metals in two spatial dimensions. II. Spin density wave order, \href{http://journals.aps.org/prb/abstract/10.1103/PhysRevB.82.075128}{Phys. Rev. B \textbf{82}, 075128}.

\bibitem{Sachdev Sokol} S. Sachdev, A. V. Chubukov, and A. Sokol, Crossover and scaling in a nearly antiferromagnetic Fermi liquid in two dimensions, \href{http://journals.aps.org/prb/abstract/10.1103/PhysRevB.51.14874}{Phys. Rev. B \textbf{51}, 14874}.

\bibitem{Sachdev Georges} S. Sachdev and A. Georges, Charge and spin-density-wave ordering transitions in strongly correlated metals, \href{http://journals.aps.org/prb/abstract/10.1103/PhysRevB.52.9520}{Phys. Rev. B \textbf{52}, 9520}.

\bibitem{Hartnoll Sachdev} S. A. Hartnoll, R. Mahajan, M. Punk, and S. Sachdev, Transport near the Ising-nematic quantum critical point of metals in two dimensions, \href{http://journals.aps.org/prb/abstract/10.1103/PhysRevB.89.155130}{Phys. Rev. B \textbf{89}, 155130}.

\bibitem{Fitzpatrick Raghu} A. L. Fitzpatrick, S. Kachru, J. Kaplan, and S. Raghu, Non-Fermi-liquid behavior of large-$N_B$ quantum critical metals, \href{http://journals.aps.org/prb/abstract/10.1103/PhysRevB.89.165114}{Phys. Rev. B \textbf{89}, 165114}.

\bibitem{Sachdev Vojta} S. Sachdev, M. Troyer, and M. Vojta, Spin Orthogonality Catastrophe in Two-Dimensional Antiferromagnets and Superconductors, \href{http://journals.aps.org/prl/abstract/10.1103/PhysRevLett.86.2617}{Phys. Rev. Lett. \textbf{86}, 2617}


\bibitem{Allais Sachdev} A. Allais and S. Sachdev, Spectral function of a localized fermion coupled to the Wilson-Fisher conformal field theory, \href{http://journals.aps.org/prb/abstract/10.1103/PhysRevB.90.035131}{Phys. Rev. B \textbf{90}, 035131}.

\bibitem{Billo Pellegrini} M. Bill\'o, M. Caselle, D. Gaiotto, F. Gliozzi, M. Meineri, R. Pellegrini, Line defects in the 3d Ising model, \href{http://arxiv.org/abs/1304.4110}{arXiv:1304.4110}

\bibitem{Pellissetto} A. Pelissetto, E. Vicari, Critical Phenomena and Renormalization Group Theory, \href{http://www.sciencedirect.com/science/article/pii/S0370157302002193}{Phys. Rep. \textbf{368}, 549}.

\bibitem{Wolff} U. Wolff, Collective Monte Carlo updating for spin systems, \href{http://journals.aps.org/prl/abstract/10.1103/PhysRevLett.62.361}{Phys. Rev. Lett. \textbf{62}, 361}.

\bibitem{Binder}K. Binder, D. P. Landau, A guide to Monte Carlo simulations in statistical physics, 2nd ed. pag. 139.
\end{document}